\documentclass{webofc}
\usepackage[varg]{txfonts}   
\begin{document}
\title{Anomalous transport phenomena and momentum space topology}
%
%

\author{\firstname{Mikhail} \lastname{Zubkov}\inst{1}\fnsep\thanks{Talk presented at the  XXth International Seminar "Quarks-2018" (27 May 2018 - 2 June 2018, Valday, Russia).}\fnsep\thanks{On leave of absence from Institute for Theoretical and Experimental Physics, B. Cheremushkinskaya 25, Moscow, 117259, Russia. \email{zubkov@itep.ru}} \and
        \firstname{Zakhar} \lastname{Khaidukov}\inst{2}\fnsep\thanks{\email{khaidukov@itep.ru}} }

\institute{Physics Department, Ariel University, Ariel 40700, Israel
\and
Institute for Theoretical and Experimental Physics, B. Cheremushkinskaya 25, Moscow, 117259, Russia}

\abstract{%
 Using the derivative expansion applied to the Wigner transform of the two - point Green
function this is possible to derive the response of various nondissipative currents to the
external gauge fields. The corresponding currents are proportional to the momentum space
topological invariants. This allows to analyse systematically various anomalous transport
phenomena including the anomalous quantum Hall effect and the chiral separation effect.
We discuss the application of this methodology both to the solid state physics and to the
high energy physics.
}
\maketitle
\section{Introduction}
\label{intro}
The so - called non - dissipative transport effects have been widely discussed recently \cite{Landsteiner:2012kd,semimetal_effects7,Gorbar:2015wya,Miransky:2015ava,Valgushev:2015pjn,Buividovich:2015ara,Buividovich:2014dha,Buividovich:2013hza}. In the high energy physics those effects are expected to be observed in the non - central heavy ion collisions, when the fireballs are in the presence of both magnetic field and rotation~\cite{ref:HIC}. Such effects have also been considered for the recently discovered Dirac and Weyl semimetals \cite{semimetal_effects6,semimetal_effects10,semimetal_effects11,semimetal_effects12,semimetal_effects13,Zyuzin:2012tv,tewary,16}.
It was expected that their family includes, in particular,  the chiral separation effect (CSE) \cite{Metl}, the chiral magnetic effect (CME)  \cite{CME,Kharzeev:2015znc,Kharzeev:2009mf,Kharzeev:2013ffa}, the chiral vortical effect (CVE) \cite{Vilenkin}, the anomalous quantum Hall effect (AQHE)  \cite{TTKN,Hall3DTI,Zyuzin:2012tv}.

It is widely believed that all mentioned above phenomena have the same origin - the chiral anomaly. For example, the direct "derivation" of the CME from the standard expression for chiral anomaly has been presented \cite{Zyuzin:2012tv,CME,SonYamamoto2012}.
Nevertheless, the more accurate consideration demonstrates that the original equilibrium version of the CME does not exist. In \cite{Valgushev:2015pjn,Buividovich:2015ara,Buividovich:2014dha,Buividovich:2013hza} this was demonstrated using the numerical lattice methods. In the context of condensed matter theory the absence of CME was reported for the particular model of Weyl semimetal \cite{nogo}. The same conclusion has been obtained using the conjectured no - go Bloch theorem \cite{nogo2}. The analytical proof of the absence of the equilibrium CME was presented in \cite{Z2016_1,Z2016_2} using the technique of Wigner transformation \cite{Wigner,star,Weyl,berezin}  applied to the lattice models\footnote{The CME may, though, survive in a certain form as a non - equilibrium kinetic phenomenon - see, for example,  \cite{ZrTe5}.}. The same technique allowed to reproduce the known results on the AQHE \cite{Z2016_1} and to confirm the existence of the equilibrium CSE \cite{KZ2017} both in the framework of the quantum field theory and the solid state physics.

Below we review the mentioned above technique, which allows to express through the topological invariants in momentum space the response of various nondissipative currents to the external magnetic or electric field.
In the framework of the naive nonregularized quantum field theory the CSE was discussed, for example, in \cite{Gorbar:2015wya} in the technique similar to the one that was used for the consideration of the CME \cite{CME} in Dirac semimetals and AQHE in Weyl semimetals \cite{Zyuzin:2012tv}.  However, the more refined calculations reported here confirm the existence of the CSE and AQHE in the mentioned systems and demonstrate the absence of the equilibrium CME \cite{Z2016_1,Z2016_2,KZ2017}.

It is worth mentioning that the investigation of momentum space topology was developed previously mainly within the condensed matter physics theory. It allows to relate the gapless boundary fermions to the bulk topology for the topological insulators and to describe the stability of the Fermi surfaces and the Fermi points, as well as the fermion zero modes on vortices (for the review see \cite{Volovik2003,Volovik:2011kg}). It is worth mentioning that momentum space topology of QCD has been discussed recently in \cite{Z2016_3}. The whole Standard Model of fundamental interaction has been considered as a topological material in \cite{Volovik:2016mre}.


\section{Lattice models in momentum space}

\label{SectWigner}

Here we consider briefly the lattice models in momentum space following the methodology of \cite{Z2016_1,Z2016_2}.
In the absence of the external gauge field  the partition function of the theory defined on the infinite lattice is
\begin{equation}
Z = \int D\bar{\psi}D\psi \, {\rm exp}\Big( - \int_{\cal M} \frac{d^D {p}}{|{\cal M}|} \bar{\psi}^T({p}){\cal G}^{-1}({ p})\psi({p}) \Big)\label{Z1}
\end{equation}
 Here $|{\cal M}|$ is the volume of momentum space $\cal M$, $D$ is the dimensionality of space - time. $\bar{\psi}$ and $\psi$ are the Grassmann - valued fields defined in momentum space $\cal M$. The Green function $\cal G$ is specific for the given system. For the often used model of $3+1$ D Wilson fermions it is given by
 \begin{equation}
 {\cal G}({p}) = - i \Big(\sum_{k}\gamma^{k} g_{k}({p}) - i m({p})\Big)^{-1}\label{G10}
 \end{equation}
 where $\gamma^k$ are Euclidean Dirac matrices
while
$g_k({p})$ and $m({p})$ are the real - valued functions ($k = 1,2,3,4$) given by
\begin{equation}
g_k({p}) = {\rm sin}\,
p_k, \quad m({p}) = m^{(0)} +
\sum_{a=1,2,3,4} (1 - {\rm cos}\, p_a)\label{gWilson}
\end{equation}
The fields in coordinate space are related to the fields in momentum space:
$\psi({r}) = \int_{\cal M} \frac{d^D {p}}{|{\cal M}|} e^{i {p}{r}} \psi({p})$.
 This allows to define formally the values of fields at any other values of $r$, not only at the lattice sites. The partition function has the form
\begin{equation}
Z = \int D\bar{\psi}D\psi \, {\rm exp}\Big( - \sum_{{r}_n} \bar{\psi}^T({r}_n)\Big[{\cal G}^{-1}(-i\partial_{r})\psi({ r})\Big]_{{r}={r}_n} \Big)\label{Z2}
\end{equation}
Here the sum in the exponent is over the discrete coordinates ${r}_n$. However, the operator $-i\partial_{r}$ acts on the function $\psi({r})$ of continuous variable defined above.

\section{Introduction of the gauge field}

In the presence of the constant external gauge field corresponding to the potential $A(x)$ (up to the terms irrelevant in the low energy effective theory) we may represent the partition function as follows \cite{Z2016_1,Z2016_2}
\begin{eqnarray}
Z = \int D\bar{\psi}D\psi \, {\rm exp}\Big( -  \int_{\cal M} \frac{d^D {p}}{|{\cal M}|} \bar{\psi}^T({p})\hat{\cal Q}(i{\partial}_{p},{p})\psi({p}) \Big)\label{Z4}
\end{eqnarray}
Here
\begin{equation}
\hat{\cal Q} = {\cal G}^{-1}({p} - {A}(i{\partial}^{}_{p}))\label{calQM}
\end{equation}
while the pseudo - differential operator ${A}(i\partial_{p})$ is defined as follows. First, we represent the original gauge field ${A}({r})$ as a series in powers of coordinates ${r}$. Next, variable ${r}$ is substituted in this expansion by the operator $i\partial_{p}$. Besides, in Eq. (\ref{calQM}) each product of the components of ${p} - {A}(i{\partial}^{}_{p})$ is subsitituted by the symmetric combination (for the details see \cite{Z2016_1}).


Electric current is the response of effective action to the variation of the external Electromagnetic field. This gives
\begin{eqnarray}
j^k({R}) &=& -\int_{\cal M} \frac{d^D {p}}{(2\pi)^D} \,  {\rm Tr} \, \tilde{G}({R},{p}) \frac{\partial}{\partial p_k}\Big[\tilde{G}^{(0)}({R},{p})\Big]^{-1}\label{j423}
\end{eqnarray}
where the Wigner transformation of the Green function is expressed as:
\begin{equation}
 \tilde{G}({R},{p}) = \sum_{{r}={r}_n} e^{-i {p} {r}} G({R}+{r}/2,{R}-{r}/2)\label{Wl2}
\end{equation}
while the Green function itself is
\begin{eqnarray}
G({r}_1,{r}_2)&=& -\frac{1}{Z}\int D\bar{\Psi}D\Psi \,\bar{\Psi}({r}_2)\Psi({r}_1) {\rm exp}\Big(-\sum_{{ r}_n}\Big[ \bar{\Psi}({r}_n)\Big[{\cal G}^{-1}(-i\partial_{r}  \nonumber\\&& - {A}({r}))\Psi({r})\Big]_{{r}={r}_n}\Big]\Big)\nonumber
\end{eqnarray}
At the same time $
\tilde G^{(0)}({R},{p})  = {\cal G}({p}-{A}({R}))\label{Q0}
$.
In \cite{Z2016_1} the following expression was derived for the linear response of the electric current to external electromagnetic field: \begin{eqnarray}
j^{(1)k}({R})  &= & - \frac{1}{4\pi^2}\epsilon^{ijkl} {\cal M}_{l} A_{ij} ({R}), \label{calM}\\
{\cal M}_l &=& \int_{} \,{\rm Tr}\, \nu_{l} \,d^4p \label{Ml} \\ \nu_{l} & = &  - \frac{i}{3!\,8\pi^2}\,\epsilon_{ijkl}\, \Big[  {\cal G} \frac{\partial {\cal G}^{-1}}{\partial p_i} \frac{\partial  {\cal G}}{\partial p_j} \frac{\partial  {\cal G}^{-1}}{\partial p_k} \Big]  \label{nuG}
\end{eqnarray}

Here tensor $\cal M$ is the topological invariant in momentum space, i.e. it is not changed if the system is modified smoothly. This representation allows to prove that the equilibrium CME does not exist for the system of massless fermions. The proof is presented in \cite{Z2016_2}. At the nonzero chiral chemical potential  $\mu_5$ we may consider the following expression for the fermion Green function in the absence of the gauge field:
 \begin{equation}
 {\cal G}^{}({\bf p}) = \Big(\sum_{k}\gamma^{k} g_{k}({\bf p}) + i\gamma^4 \gamma^5 \mu_5 - i m({\bf p})\Big)^{-1}\label{G2}
 \end{equation}
 where $g_k({\bf p})$ and $m({\bf p})$ are the real - valued functions, $k = 1,2,3,4$. Function $m$ is defined in such a way that the pole of $\cal G$ appears that corresponds to the massless fermion.
We may substitute ${\cal G}$ of Eq. (\ref{G2}) into Eq. (\ref{calM}) while dealing with the linear response to the external magnetic field. In the non - marginal cases including the ordinary regularization of the QFT using Wilson fermions we are able to bring the system using smooth transformation from the state with nonzero $\mu_5$ and vanishing fermion mass to the state with vanishing $\mu_5$ and nonzero fermion mass. During such a modification the singularity of the Green function is not encountered. Therefore, the value of the topological invariant ${\cal M}_4$ responsible for the CME is not changed. It may be calculated easily for the system of massive fermions, and this calculation gives ${\cal M}_4 =0$, which proves the absence of the corresponding CME current.

\section{AQHE in the $3+1$D systems}
\label{SectHall3d}


From the above expressions it follows that the Hall current is given by \cite{Z2016_1}
\begin{equation}
{j}^k_{Hall} = -\frac{1}{4\pi^2}\,{\cal M}^\prime_l\,\epsilon^{jkl}E_j,\label{HALLj3d}
\end{equation}
where ${\cal M}^\prime_l = i{\cal M}_l/2$ is
\begin{eqnarray}
{\cal M}^\prime_l &=&  \frac{1}{3!\,4\pi^2}\,\epsilon_{ijkl}\,\int_{} \,\,d^4p\,{\rm Tr} \Big[  {\cal G} \frac{\partial {\cal G}^{-1}}{\partial p_i} \frac{\partial  {\cal G}}{\partial p_j} \frac{\partial  {\cal G}^{-1}}{\partial p_k} \Big]  \label{nuGHall}
\end{eqnarray}
Notice, that in the case of the non - interacting condensed matter system with the Green function  ${\cal G}^{-1} = i \omega - \hat{H}({\bf p})$ expressed through the Hamiltonian $\hat{H}$ we may derive (similar to the above considered case of the $2+1$ D system) the representation of the components of topological invariant ${\cal M}^{\prime}_l$ with $l\ne 4$ through the Berry curvature $ {\cal F}_{ij}$:
\begin{eqnarray}
{\cal M}^\prime_l &=&  \frac{\epsilon^{ijl}}{4\pi}\sum_{\rm occupied}\, \int d^3p\, {\cal F}_{ij}
\end{eqnarray}
Here the sum is over the occupied branches of spectrum.

The formalism developed in \cite{Z2016_1} allows to demonstrate that for the wide class of Weyl semimetals the contribution to the AQHE electric current of the pair of Weyl fermions is given by
\begin{equation}
{j}^k_{Hall} = \frac{\beta}{2\pi^2}\,\epsilon^{jk3}E_j,\label{HALLj3dp}
\end{equation}
Here $E$ is electric field while $\beta$ is the distance in momentum space between the Weyl fermions of opposite chirality while the third axis is directed along the line connecting them. Thus the previously obtained result on the AQHE in Weyl semimetals is confirmed. Moreover, the same method allows to predict the existence of the topological insulators with the AQHE, in which the meaning of the constant $\beta$ is the length of the inverse lattice vector (proportional to $1/a$, where $a$ is the lattice spacing).

\section{Linear response of axial current to external magnetic field}

In continuum theory the naive expression for the axial current is $\langle \bar{\psi} \gamma^\mu \gamma^5 \psi\rangle$. Several different definitions for the particular lattice regularization may give this expression in the naive continuum limit. In \cite{KZ2017} it has been proposed to define the axial current in the lattice models as follows
\begin{eqnarray}
j^{5k}({R}) &=& \int_{\cal M} \frac{d^D {p}}{(2\pi)^D} \,  {\rm Tr} \,\gamma^5\,  \tilde{G}({R},{p}) \frac{\partial}{\partial p_k}\Big[\tilde{G}^{(0)}({R},{p})\Big]^{-1}\label{j423}
\end{eqnarray}
where
\begin{eqnarray}
&&\tilde G^{(0)}({R},{p})  = {\cal G}({p}-{A}({R}))\label{Q01}
\end{eqnarray}
One can easily check that in the naive continuum limit this definition gives $\langle \bar{\psi} \gamma^k \gamma^5 \psi\rangle$.

Let us regularize the expressions for the massless fermions using the finite temperature version of the lattice theory. For the periodic boundary conditions in the spatial directions
and anti-periodic in the imaginary time direction, the lattice momenta are
\begin{equation}\label{disc}
p_i \in (0, 2\pi);\, p_4=\frac{2\pi}{N_t }(n_4+1/2)
\end{equation}
where $i=1,2,3$ while $n_4=0,...,N_t-1$. Temperature is equal to $T = 1/N_t $, in lattice units $1/a$, where $a$ is the lattice spacing. Thus the imaginary frequencies are discrete $p_4 = \omega_{n}=2\pi T (n+1/2)$, where $n= 0, 1, ... N_t-1$, while the axial current is expressed via the Green's functions as follows:
\begin{eqnarray}
j^{5k}&=&-\frac{i}{2}T\sum_{n=0}^{N_t-1}\int \frac{d^3p}{(2\pi)^3}{\rm Tr}\, \gamma^5 ({\cal G}(\omega_{n},\textbf{p})\partial_{p_{i}}{\cal G}^{-1}(\omega_{n},\textbf{p})\nonumber \\&&\partial_{p_{j}}{\cal G}(\omega_{n},\textbf{p})\partial_{p_{k}}{\cal G}^{-1}(\omega_{n},\textbf{p}))F_{ij}\label{calN}
\end{eqnarray}


Next, we introduce the chemical potential in the standard way $\omega_{n} \to \omega_{n}-i\mu$. If $\gamma^5$ anti - commutes with the Green function in a small vicinity of its poles, then the term in the axial current linear in  $\mu$ is \cite{KZ2017}
\begin{equation}
j^{5k}= \frac{{\cal N}\,\epsilon^{ijk}}{4\pi^2} F_{ij} \mu\label{jmuH}
\end{equation}
in the low temperature limit $T\to 0$. Here
\begin{eqnarray}
{\cal N}&=&\frac{\epsilon_{ijkl}}{12}\int_{\Sigma}\frac{d \sigma^l}{(2\pi)^2}{\rm Tr}\, \gamma^5 {\cal G}(\omega_{},\textbf{p})\partial_{p_{i}}{\cal G}^{-1}(\omega_{},\textbf{p})\nonumber \\&&\partial_{p_{j}}{\cal G}(\omega_{},\textbf{p}) \partial_{p_{k}}{\cal G}^{-1}(\omega_{},\textbf{p})\label{calN1}
\end{eqnarray}
while $\Sigma$ is the 3D hypersurface of infinitely small volume that embraces the singularities of the Green function concentrated at the Fermi surfaces (or Fermi points). The advantage of this representation is that Eq. (\ref{calN}) is the topological invariant.

In particular, in \cite{KZ2017} it has been demonstrated using the regularization with Wilson and overlap fermions that if the lattice model describes one massless Dirac fermion, then
\begin{equation}
{\cal N}= 1
\end{equation}
In this way the validity of the obtained earlier naive result for the CSE has been established.

In the presence of nonzero mass $m^{(0)}$ the situation is changed. At $\mu \ge m^{(0)}$ the Fermi surface appears, and it contributes to the chiral current through Eq. (\ref{calN}). However, the result is not given by the simple expression of Eq. (\ref{jmuH}).

\section{Conclusions}
\label{sec-1}

Above we reviewed the application of momentum space topology to the analysis of anomalous transport. This methodology works equally well both in the lattice regularized relativistic quantum field theory and for the lattice models of solid state physics. In particular, it appears that the response of electric current to the external electric field is expressed through the topological invariant in momentum space. Its nonzero value leads to the appearance of the quantum Hall effect. This allows to calculate the AQHE conductivity for the wide class of the condensed matter systems. The same technique allows to prove the absence of the equilibrium CME in the lattice regularized quantum field theory: the corresponding conductivity is proportional to the topological invariant in momentum space that vanishes for the systems with finite chiral chemical potential. For the systems of massless fermions the further development of this technique allows to express through the topological invariant the terms in the axial current proportional to the external magnetic field and to the ordinary chemical potential. This is the topological invariant in momentum space responsible for the stability of the Fermi point. In this way we describe the axial current of the CSE that is relevant for description of the fireballs that appear in the non - central heavy ion collisions.

We show that in the considered cases the corresponding nondissipative currents are proportional to the momentum space topological invariants. They are not changed when the system is deformed smoothly, without passing through a phase transition. The corresponding conductivities for the  complicated systems may be calculated within the simple ones related to the original systems by a smooth deformation. On the technical side our methodology is based on the derivative expansion applied to the Wigner transform of the two - point Green functions. We introduce the slowly varying external gauge field directly to the momentum space formulation of the lattice models. The external gauge field appears as a pseudo - differential operator ${\bf A}(i\partial_{\bf p})$. This way of the incorporation of the external field to the theory is  useful for the analytical derivations and allows to obtain the above mentioned remarkable relation between the momentum space topological invariants and the non - dissipative currents.

We started our consideration from Eq. (\ref{Z1}) thus neglecting the contributions of interactions. However, even the complicated interacting system may be described in a certain approximation by the same Eq. (\ref{Z1}). Then function $\cal G$ incorporates all contributions of interactions to the two - point fermion Green functions. In this approach we neglect the contributions to the physical observables of the fermion Green functions with more than two external lines.

MZ kindly acknowledges numerous discussions with G.E.Volovik. Both authors are greatful to M.N.Chernodub for useful discussions. The work of ZK was supported by Russian Science Foundation Grant No 16-12-10059.

\end{document}